\pdfoutput=1
\documentclass[twocolumn, prb, showpacs, preprintnumbers,amsmath,amssymb, english,notitlepage,superscriptaddress,showpacs]{revtex4-1}% Physical Review B

\DeclareMathAlphabet{\mathcalligra}{T1}{calligra}{m}{n}

\usepackage{graphicx}
\usepackage[colorlinks=true,urlcolor=black,linkcolor=black,citecolor=black]{hyperref}
\usepackage{amsmath}
\usepackage{amssymb}
\usepackage{bm}% bold math

\newcommand{\beq}{\begin{equation}}
\newcommand{\eeq}{\end{equation}}
\newcommand{\bdis}{\begin{displaymath}}
\newcommand{\edis}{\end{displaymath}}
\newcommand{\bea}{\begin{eqnarray}}
\newcommand{\eea}{\end{eqnarray}}
\newcommand{\barr}{\begin{array}}
\newcommand{\earr}{\end{array}}
\newcommand{\bfig}{\begin{figure}[!]}
\newcommand{\efig}{\end{figure}}

\begin{document}

\title{Thermodynamic transport theory of spin waves in ferromagnetic insulators}

\author{Vittorio Basso}
\affiliation{Istituto Nazionale di Ricerca Metrologica, Strada delle Cacce 91, 10135 Torino, Italy}
\author{Elena Ferraro}
\email{Present address: Laboratorio MDM, IMM-CNR, Via Olivetti 2, 20864 Agrate Brianza (MB), Italy; elena.ferraro@mdm.imm.cnr.it}
\affiliation{Istituto Nazionale di Ricerca Metrologica, Strada delle Cacce 91, 10135 Torino, Italy}
\author{Marco Piazzi}
\affiliation{Istituto Nazionale di Ricerca Metrologica, Strada delle Cacce 91, 10135 Torino, Italy}

\date{\today}

\begin{abstract}
We use the Boltzmann transport theory in the relaxation time approximation to describe the thermal transport of spin waves in a ferromagnet. By treating spin waves as magnon excitations we are able to compute analytically and numerically the coefficients of the constitutive thermo-magnetic transport equations. As a main result, we find that the absolute thermo-magnetic power coefficient $\epsilon_M$, relating the gradient of the potential of the magnetization current and the gradient of the temperature, in the limit of low temperature and low field, is a constant $\epsilon_M = -0.6419 \, k_B/\mu_B$. The theory correctly describes the low-temperature and magnetic-field dependencies of spin Seebeck experiments. Furthermore, the theory predicts that in the limit of very low temperatures the spin Peltier coefficient $\Pi_M$, relating the heat and the magnetization currents, tends to a finite value which depends on the amplitude of the magnetic field. This indicates the possibility to exploit the spin Peltier effect as an efficient cooling mechanism in cryogenics.
\end{abstract}


\maketitle
%\tableofcontents

%%%%%%%%%%%
%%%%%%%%%%%%
%%%%%%%%%%%%%

%%%%%%%%%%%%%%%%%%%%%%%%%%%%%%%%%%%%%%%
%%%%%%%%%%%%%%%%%%%%%%%%%%%%%%%%%%%%%%%
%%%%%%%%%%%%%%%%%%%%%%%%%%%%%%%%%%%%%%%
\section{Introduction}\label{SEC:Intro}

The thermal transport of spin waves is the key physical phenomenon at the basis of the longitudinal spin Seebeck effect (LSSE) \cite{Uchida-2010}, which is one of the most promising effect of spin caloritronics \cite{Zutic-2004, Bauer-2012, Boona-2014}. In the LSSE a spin current is generated by a temperature gradient in a ferrimagnetic insulator like yttrium iron garnet (YIG) and it is detected by the inverse spin Hall effect in an adjacent Pt layer \cite{Uchida-2010c, Uchida-2012, Uchida-2013b, Kikkawa-2013, Miao-2013}. This effect has been interpreted as the consequence of the presence of a spin current carried by non equilibrium spin waves (or magnons) in the ferrimagnetic insulator which is converted into a spin current carried by electrons in the metal \cite{Bauer-2012, Adachi-2013, Siegel-2014}. The source of this conversion is the interaction between the $d$ electrons in the ferromagnet and the $s$ electrons in the metal which gives rise to an absorption of a magnon in the insulator and the corresponding generation of an electron spin flip in the metal \cite{Brataas-2000, Ohe-2011, Qiu-2013, Qiu-2013, Hoffmann-2013, Tveten-2015, Chotorlishvili-2015, Bender-2015}. While the assessment of the interface is well documented in the literature \cite{Xiao-2010, Xiao-2010b, Adachi-2013, Schreier-2013}, the description of the transport of non equilibrium spin waves inside the bulk ferromagnet is still a debated issue \cite{Zhang-2012, Zhang-2012b, Rezende-2014, Rezende-2014b, Jin-2015, Nakata-2015b, Rezende-2016, Basso-2016}. 

The thermodynamic problem of the joint transport of heat and magnetic moment (thermo-magnetics) has its counterpart in the field of thermoelectricity where one has the joint transport of heat and charge. In the non equilibrium thermodynamics of fluxes and forces these joint transport problems are described by the identification of the current densities of the extensive variables of interest with their associated thermodynamic forces. While in the field of thermoelectricity it is well assessed that the gradient of the electric potential $\nabla V_e$ is the force associated to the electric current density $\boldsymbol{j}_e$ and the temperature gradient $\nabla T$ is the force associated to the heat current $\boldsymbol{j}_q$, in the field of thermo-magnetics there is controversy regarding the thermodynamic forces to employ. The reason is related to the correct determination of the relations between the thermodynamic description and the microscopic details of the processes.

For what concerns the temperature variable, while for static magnetic properties the thermodynamic temperature is assumed to be the correct one \cite{Anderson-1964}, for the transport properties the question has not an obvious answer. The problem dates back to the 1977 paper of Sanders and Walton \cite{Sanders-1977} discussing the reasons for the elusive contribution of magnons to the total heat conductivity of magnetic materials. Sanders and Walton noted that the magnon subsystem could be partially isolated from the phonon one and described the equilibration of the temperatures of the two subsystems by means of the finite relaxation time $\tau_{mp}$. In the limit of very long relaxation times, the two temperatures are different and only the phonon heat transport could be experimentally observed, while, if $\tau_{mp}$ is short enough, the two temperatures coincide with the thermodynamic temperature and the magnon heat transport can be observed. When, more recently, this picture was applied to the spin Seebeck effect, it turned out that, assuming partial decoupling, most of the transport effects could just be described as proportional to the difference between the temperature of the magnons in the ferromagnet and the temperature of the electrons in the metal \cite{Xiao-2010, Xiao-2010b, Adachi-2013, Schreier-2013}. This approach has been successful to describe the proportionality of the injected spin current on the temperature gradient, but it was unable to correctly reproduce the experimental dependence on the thickness of YIG \cite{Kehlberger-2015}, therefore showing the necessity to take into account also the bulk and not only the interface. Stimulated also by the experimental results of Ref.[\onlinecite{Agrawal-2013}], showing that in YIG there exists a close spatial correspondence between magnons and phonons temperatures, other authors have used the gradient of the thermodynamic temperature as a driving force \cite{Zhang-2012, Zhang-2012b, Rezende-2014, Rezende-2014b, Jin-2015, Rezende-2016}, therefore assuming a good thermal coupling between the two subsystems. It is worth to mention that Sanders and Walton \cite{Sanders-1977} in their original paper already demonstrated that, for YIG, magnons and phonons could exchange energy with a quite short time constant $\tau_{mp}$. However, as $\tau_{mp}$ may be material dependent and not too small, the general case, which gives rise to non trivial coupled phonon-magnon heat flow, is indeed of interest and it has been treated in Ref.[\onlinecite{Liao-2014}]. In the limit of very small $\tau_{mp}$ the picture simplifies and the phonon and magnon heat flows can be just linearly summed up.

Much more subtle is the choice of the thermodynamic force to be associated with the magnetic moment current. Refs.[\onlinecite{Zhang-2012, Zhang-2012b, Rezende-2014, Rezende-2014b, Rezende-2016}] have used the gradient of the magnon accumulation as a force. This approach, while microscopically correct, suffers for the problem that, in non equilibrium thermodynamics, the current of an extensive variable should be always associated to the gradient of an intensive variable. This effective force has been worked out in the thermodynamic theory of Johnson and Silsbee \cite{Johnson-1987}. In their approach the potential for the magnetization current is derived from purely thermodynamic grounds and it results to be given by $H^* = H-H_{eq}$, the difference between the magnetic field $H$ and the field $H_{eq}(M)$ corresponding to the equation of state at equilibrium \cite{Basso-2016}. When the theory of Johnson and Silsbee is applied to the problem of the transport of magnons, one finds that the potential $H^*$ is directly related to the chemical potential $\mu_m$ associated with the magnon accumulation, that is $H^* =  \mu_m/(2 \mu_0\mu_B)$. It is worth to notice that the gradient of the chemical potential of magnons is also used in the very recent publication of Ref.[\onlinecite{Cornelissen-2016}].

In the present paper we apply the Boltzmann transport equation to magnons by using the gradient of the Johnson and Silsbee potential $\nabla H^*$ as the driving force for the magnetization current and the gradient of the thermodynamic temperature $\nabla T$ as the driving force for the heat current. By performing the calculations in the relaxation time approximation, we are able to derive all the kinetic coefficients of the constitutive equations. In particular we are able to obtain the expressions for the temperature and magnetic field dependences of the kinetic coefficients. These expressions have closed analytical forms in the low temperature limit, where the quadratic dispersion relation for magnons holds, and they can be computed numerically, by assuming isotropic behavior, in the high temperature range. 

As a main result we find that the absolute thermo-magnetic power coefficient $\epsilon_M$, relating $\nabla H^*$ and $\nabla T$, is independent of the magnon scattering mechanism and its value, in the limit of low field, is the constant $\epsilon_M = -0.6419 \, k_B/\mu_B$. Moreover, we predict that the spin Peltier coefficient $\Pi_M$, relating the heat and the magnetization currents, tends, at very low temperatures, to the finite value $\Pi_M = -\mu_0H_0$, which depends on the amplitude of the magnetic field $H_0$. This feature indicates the possibility to exploit the spin Peltier effect as an efficient cooling mechanism in the cryogenic temperature range. Finally we make a preliminary test of the model by inserting the parameters of YIG and computing the values of the spin Seebeck effect. By comparing with experiments at low temperature, we find that the theory correctly describes both the temperature and the magnetic field dependences of the spin Seebeck coefficient for the bulk case \cite{Kikkawa-2015}.

The paper is organized as follows. Section~\ref{SEC:transport_eqs} presents the main features of the Johnson and Silsbee theory. Section~\ref{SEC:statistics_magnons} introduces the spin waves in equilibrium under a non zero chemical potential. The Boltzmann transport equation is described in Section~\ref{SEC:transport_magnons} and the main results are presented in Section~\ref{SEC:results}. The application of the theory to the spin Seebeck effect is described in Section~\ref{SEC:SSE}. Finally, the features and the limitations of the theory are shortly discussed in Section~\ref{SEC:conclusions}. All the technical and mathematical aspects are detailed in the Appendices~\ref{APPENDIX_A} and \ref{APPENDIX_B}.

%%%%%%%%%%%%%%%%%%%%%%%%%%%%%%%%%%%%%%%
%%%%%%%%%%%%%%%%%%%%%%%%%%%%%%%%%%%%%%%
%%%%%%%%%%%%%%%%%%%%%%%%%%%%%%%%%%%%%%%
\section{Thermo-magnetic transport equations}\label{SEC:transport_eqs}

We consider a magnetic system in which the macroscopic magnetization vector $\boldsymbol{M}$ and the magnetic field $\boldsymbol{H}$ are parallel and directed along an easy axis that can be due, for example, to an anisotropic crystal structure. Therefore we describe the system by the scalar amplitudes $M$ and $H$. By following the Johnson and Silsbee approach \cite{Johnson-1987, Basso-2016}, the thermodynamics of the ferromagnet is formulated by expressing the variation of the non equilibrium system enthalpy $u_e(s,H;M)$ as a function of the entropy density $s$, of the magnetic field $H$ and of the magnetization $M$:

\begin{equation}
du_e = Tds - \mu_0MdH  - \mu_0 H^*dM.
\label{EQ:du_e}
\end{equation}

\noindent In the previous expression $T$ is the thermodynamic temperature, $\mu_0$ is the permeability of vacuum and $H^* = H-H_{eq}$ is the affinity associated to the change of the magnetization, with $H_{eq}(M)$ being the equation of state for the magnetic field at equilibrium. $H^*$ is different from zero when the system is locally out-of-equilibrium. Relaxation toward equilibrium is described by assuming linear system behavior. The relaxation equation is given by the proportionality between the rate of change of the magnetization $dM/dt$, a generalized velocity, and $H^*$, a generalized force. By generalizing this treatment to spatially extended systems in which magnetization currents $\boldsymbol{j}_M$ may be present, the relaxation equation becomes the continuity equation for the magnetization

\begin{equation}
\label{EQ:cont}
\frac{\partial M}{\partial t} + \nabla \cdot \boldsymbol{j}_M = \frac{H^*}{\tau_M}.
\end{equation}

\noindent The right hand side of Eq.(\ref{EQ:cont}) expresses the presence of sources and sinks in the system, giving rise to the non conservation of the magnetization, and it contains a phenomenological time constant $\tau_M$. Beyond its continuity equation, every medium is also characterized by its constitutive equations describing how the magnetization and the heat currents depend on the gradients of the associated intensive variables. Within this approach the relevant gradients are $\nabla T$, which is associated to the heat current $\boldsymbol{j}_q$, and $\nabla H^*$, which is associated to the magnetization current $\boldsymbol{j}_M$. By considering linear behavior and limiting to one dimension, the equations are

\begin{eqnarray}
j_M  &=& \sigma_M \, \mu_0 \nabla H^* - \epsilon_M \sigma_M \, \nabla T \label{EQ:M_curr}\\
j_q  &=& \epsilon_M \sigma_M T \mu_0 \nabla H^* - (\kappa + \epsilon_M^2 \sigma_M T) \nabla T.
\label{EQ:q_curr}
\end{eqnarray}

\noindent In Eqs.(\ref{EQ:M_curr}) and (\ref{EQ:q_curr}) $\nabla \equiv \partial/\partial x$ and, in analogy with the thermoelectric effects \cite{Callen-1985, Solyom-2007}, $\sigma_M$ is the conductivity for the magnetization current, $\epsilon_M$ is the absolute thermo-magnetic power coefficient and $\kappa$ is the total thermal conductivity under zero magnetization current. When the constitutive equation (\ref{EQ:M_curr}) is solved together with the continuity equation (\ref{EQ:cont}), under given boundary conditions, the currents and the potential $H^*$ can be computed and the typical diffusion length, $l_M = (\mu_0 \sigma_M \tau_M)^{1/2}$, results to be related to material dependent parameters \cite{Basso-2016}. In the spin Seebeck effect, in which the ferromagnetic material is subjected to a uniform temperature gradient $\nabla T$, the magnetization current source is $j_{MS} = - \sigma_M \epsilon_M  \, \nabla T$. The part of the current transmitted to the sensing layer depends on the coupling between the two materials and in particular it depends on the ratio $v_M = l_M/\tau_M$ between the diffusion length and the time constant of each layer \cite{Basso-2016}. 

All the parameters contained in the continuity equation (\ref{EQ:cont}) and the constitutive equations (\ref{EQ:M_curr}) and (\ref{EQ:q_curr}) can be calculated on the basis of a microscopic theory. In Sections~\ref{SEC:statistics_magnons} and \ref{SEC:transport_magnons} we will compute them in the case of the magnons in a ferromagnet. For what concerns the thermal conductivity we will limit to compute the contribution from spin waves. In fact, as we assume that the magnetic subsystem is well thermally coupled with all the other degrees-of-freedom, the thermodynamic temperature is the only thermal intensive variable associated to the heat current. If there are many different types of carriers, i.e. magnons and phonons, the total thermal conductivity is just the sum of the individual contributions. Instead, when the magnetic subsystem is partially decoupled from the lattice, alternative methods, as those used in Ref.[\onlinecite{Liao-2014}], can be applied.

%%%%%%%%%%%%%%%%%%%%%%%%%%%%%%%%%%%%%%%
%%%%%%%%%%%%%%%%%%%%%%%%%%%%%%%%%%%%%%%
%%%%%%%%%%%%%%%%%%%%%%%%%%%%%%%%%%%%%%%
\section{Statistics of magnons}\label{SEC:statistics_magnons}

%%%%%%%%%%%%%%%%%%%%%%%%%%%%%%%%%%%%%%%
%%%%%%%%%%%%%%%%%%%%%%%%%%%%%%%%%%%%%%%
\subsection{Spin waves}\label{SUBSEC:spin_waves}

In a saturated ferromagnet, the local magnetic moments are all essentially parallel and the spin waves are the lowest energy excitations. 

In a classical picture, the spin waves are plane waves formed by the small deviations of the magnetization vector $\boldsymbol{M}_{\bot}$ from its main direction $\boldsymbol{M}_{\|}$. Due to the precession of magnetization around the magnetic field, $\boldsymbol{M}_{\bot}$ is a vector rotating in time in the plane perpendicular to $\boldsymbol{M}_{\|}$ with an angular velocity $\omega_{\boldsymbol{q}}$ depending on the wavenumber $\boldsymbol{q}$ of the plane wave. Considering the relevant energy terms for a ferromagnet, the dispersion relation reads $\omega_{\boldsymbol{q}} = -\gamma\mu_0[(l_{ex}^2 q^2+H_0)(l_{ex}^2 q^2+H_0+M_s\sin^2\theta_q)]^{1/2}$ where $\gamma = -e/m_e$ is the gyromagnetic ratio for electrons, $l_{ex}$ is the exchange length, $H_{0}= H_a + H_{AN}$ contains here contributions from the applied field $H_a$ and the anisotropy field $H_{AN}$ along the easy axis, $M_s$ is the saturation magnetization, $q=|\boldsymbol{q}|$ and $\theta_q$ is the angle formed by the $\boldsymbol{q}$ vector with the magnetization axis \cite{Gurevich-1996, Stancil-2009}. The exchange length is related to the exchange stiffness $A$ by the relation $l_{ex}^2 = 2A/(\mu_0M_s^2)$. 

In a quantum picture, the magnon is the quantum of the spin wave, it has energy $\epsilon(\boldsymbol{q}) = \hslash \omega_{\boldsymbol{q}}$ and, in the case of magnetism due to the electron spin, it carries a magnetic moment of $-2\mu_B$, where $\mu_B = e\hslash/(2m_e)$ is the Bohr magneton. Having an integer spin, magnons obey the Bose-Einstein statistics described by the distribution

\beq
g_0(\boldsymbol{q}) = \frac{1}{ \exp\left(\epsilon(\boldsymbol{q})/(k_B T)\right)-1},
\label{EQ:BEEQ}
\eeq

\noindent where $k_B$ is the Boltzmann constant. All the relevant informations on the system are contained in the magnon energy dispersion relation $\epsilon(\boldsymbol{q})$. At low $q$ it is possible to approximate the dispersion relation as

\begin{equation}
\epsilon(\boldsymbol{q}) \simeq D q^2 + \epsilon_H,
\label{EQ:dispersion_lq}
\end{equation}

\noindent where $D = 4Aa^3$ is the stiffness constant of the spin waves, with $a$ being the interatomic distance \cite{Solyom-2007}, and $\epsilon_H = 2 \mu_0\mu_B H_0$ \cite{Stancil-2009}. Eq.(\ref{EQ:dispersion_lq}) is valid when the magnetostatic energy term can be disregarded with respect to the other terms. The magnon energy is expected to show deviations from Eq.(\ref{EQ:dispersion_lq}) at large $q$ when the wavelength approaches the interatomic distance. Specific deviations depend on the type of crystal lattice \cite{Solyom-2007}. An isotropic extension of the dispersion relation for large $q$ is considered in Appendix~\ref{APPENDIX_A}. 

%%%%%%%%%%%%%%%%%%%%%%%%%%%%%%%%
%%%%%%%%%%%%%%%%%%%%%%%%%%%%%%%%
\subsection{Equilibrium thermal magnons}\label{SUBSEC:eq_thermal_magnons}

The equilibrium properties of a ferromagnet, like the magnetization and the specific heat capacity, can be directly obtained from the statistics of magnons \cite{Kittel-2005, Gurevich-1996, Solyom-2007}. The method consists in evaluating the statistical average of the corresponding physical quantities (the magnetic moment, the energy) weighted over the distribution given by Eq.(\ref{EQ:BEEQ}) and integrated over the space $\Sigma$ of the admissible wavenumbers $\boldsymbol{q}$. For example, the average magnetization is given by the integral

\begin{equation}
\label{EQ:M_def}
M = M_0 - 2\mu_B \frac{1}{(2\pi)^3}\int_{\Sigma} \, g_0(\boldsymbol{q}) d^3q,
\end{equation}

\noindent where $M_0$ is the spontaneous magnetization at zero temperature given by $M_0 = n \mu_B$, with $n$ being the density of elementary magnetic moments. The integration is performed by standard methods (see Appendix~\ref{APPENDIX_A}) and the result can be written as  

\beq
\label{EQ:M_eq}
M = M_0 \left( 1- \frac{1}{2\pi^2} t^{3/2} A_{0}\right),
\eeq

\noindent where $t=T/T_m$ is the dimensionless temperature, with $T_m ={D}/({a^2k_B})$ giving the temperature scale of the problem. By following an equivalent method for the specific heat $c_m=\partial u_m /\partial T$ the result is 

\beq
\label{EQ:cm_eq}
c_m = nk_B \frac{5}{8} \frac{1}{\pi^2}t^{3/2} A_{1}.
\eeq

\noindent Eqs.(\ref{EQ:M_eq}) and (\ref{EQ:cm_eq}) contain the well know result that the low temperature behavior of $M$ and $c_m$ is a $T^{3/2}$ power law. Furthermore additional dependence on the temperature and field is contained in the dimensionless coefficients $A_{r}$ $(r=0,1,\ldots)$, which contain a dependence on the ratio $h/t$, where $h= {2\mu_0\mu_BH_0}/({k_BT_m})$ is the dimensionless magnetic field. The definition of the coefficients $A_{r}$ is reported in Appendix~\ref{APPENDIX_A}, Eq.(\ref{EQ:Ar_def}).

%%%%%%%%%%%%%%%%%%%%%%%%%%%%%%%%
%%%%%%%%%%%%%%%%%%%%%%%%%%%%%%%%
\subsection{Relaxation of thermal magnons}\label{SUBSEC:relaxation_thermal_magnons}

When the number of magnons is different from the equilibrium value of Section~\ref{SUBSEC:spin_waves} (see Eq.(\ref{EQ:BEEQ})), one describes their distribution as 

\beq
g_0(\boldsymbol{q}, \mu_m) = \frac{1}{ \exp\left((\epsilon(\boldsymbol{q})-\mu_m)/(k_B T)\right)-1},
\label{EQ:BEQE}
\eeq

\noindent by introducing a non zero chemical potential $\mu_m$. Considering that the magnetization can change only because the number of magnons $n_m$ changes, $dM = - 2\mu_B dn_m$, by using Eq.(\ref{EQ:du_e}) with $dH=0$, one finds that $\mu_m = \partial u_e/\partial n_m = 2 \mu_0 \mu_B H^*$. This shows that for an insulating ferromagnet the chemical potential associated with the magnons, $\mu_m$, is equivalent to the Johnson and Silsbee potential $H^*$. Every time the potential $H^*$ is different from zero, the distribution of magnons is given by Eq.(\ref{EQ:BEQE}). The equilibrium distribution $g_{0}(\boldsymbol{q}, 0)$ of Eq.(\ref{EQ:BEEQ}) is recovered when $H^*=0$ \cite{Cherepanov-1993}. To describe how the distribution $g_0(\boldsymbol{q}, \mu_m)$ of Eq.(\ref{EQ:BEQE}) relaxes to the distribution $g_0(\boldsymbol{q}, 0)$ of Eq.(\ref{EQ:BEEQ}), one has to state a relaxation equation. As the time dependence can be only in the chemical potential, i.e. $\mu=\mu_m(t)$, a reasonable assumption is to consider the relaxation rate toward equilibrium to be proportional to the distance from equilibrium itself. Therefore the equation for $\mu_m(t)$ is $ {\partial \mu_m}/{\partial t}  = - {\mu_m}/{\tau_r}$, where $\tau_r$ is the magnon relaxation time constant describing the generation and annihilation of magnons. The relaxation equation for the distribution $g_0(\boldsymbol{q}, \mu_m)$ is then

\begin{equation}
\label{EQ:f0rate}
\frac{\partial g_0(\boldsymbol{q}, \mu_m)}{\partial t}  = - \frac{g_0(\mu_m) - g_{0}(0)}{\tau_r}.
\end{equation}

\noindent For a spatially homogeneous system the continuity equation (\ref{EQ:cont}), with $\boldsymbol{j}_M=0$, becomes simply $\partial M/\partial t = H^*/\tau_M$. As from Eq.(\ref{EQ:f0rate}) it is possible to derive the same equation, we are able to establish a relation between the macroscopic and the microscopic relaxation times, $\tau_M$ and $\tau_r$ respectively. On the right hand side of Eq.(\ref{EQ:f0rate}), if $g_0(\mu_m)$ is very close to $g_{0}(0)$ we can express their difference as $g_0(\mu_m)-g_{0}(0) \simeq - \mu_m \partial g_0(0)/\partial \epsilon$ and Eq.(\ref{EQ:f0rate}) becomes

\begin{equation}
\label{EQ:f0rate2}
\frac{\partial g_0(\boldsymbol{q}, \mu_m)}{\partial t}  = \frac{\mu_m}{\tau_r} \frac{\partial g_0(\boldsymbol{q}, 0)}{\partial \epsilon}.
\end{equation}

\noindent By performing the statistical average of both sides of Eq.(\ref{EQ:f0rate2}) and multiplying by $-2\mu_B$, we obtain $\partial M/\partial t $ at the left hand side. Then by using $\partial M/\partial t = H^*/\tau_M$ we obtain the expression for $\tau_M$ as 

\begin{equation}
\label{EQ:tauMdef}
\frac{1}{\tau_M}  =- \mu_0 (2 \mu_B)^2 \frac{1}{(2\pi)^3}\int_{\Sigma} \frac{1}{\tau_{r}} \frac{\partial g_{0}(0)}{\partial \epsilon} d^3q,
\end{equation}

\noindent which is the sought relation. By comparing Eq.(\ref{EQ:tauMdef}) with the definition of the magnetization of Eq.(\ref{EQ:M_eq}), it turns out the additional relation $\tau_M = \tau_r/\chi$, where $\chi = \partial M/\partial H_0$. The expression for the time constant $\tau_M$ of Eq.(\ref{EQ:tauMdef}) can be integrated (see Appendix~\ref{APPENDIX_A}), giving as a result

\begin{equation}
\tau_M = \frac{\tau_r}{m}2\pi^2 t^{-1/2}(-A_0^{\prime})^{-1},
\label{EQ:tau_Mcalc}
\end{equation}

\noindent where $m= {2\mu_0\mu_BM_0}/({k_BT_m})$ is a dimensionless parameter depending on the spontaneous magnetization $M_0$ and $A_0^{\prime}$ is a dimensionless coefficient defined in Appendix~\ref{APPENDIX_A}, Eq.(\ref{EQ:A01_def}). 

%%%%%%%%%%%%%%%%%%%%%%%%%%%%%%%%
%%%%%%%%%%%%%%%%%%%%%%%%%%%%%%%%
%%%%%%%%%%%%%%%%%%%%%%%%%%%%%%%%
\section{Transport of non equilibrium magnons}\label{SEC:transport_magnons}

%%%%%%%%%%%%%%%%%%%%%%%%%%%%%%%%%%%%%%%
%%%%%%%%%%%%%%%%%%%%%%%%%%%%%%%%%%%%%%%
\subsection{Boltzmann transport equation}\label{SUBSEC:Boltz_transp_eq}

On the basis of what shown in Section~\ref{SEC:statistics_magnons} we can describe the accumulation of magnons by means of the chemical potential $\mu_m$. The non equilibrium distribution $g(\boldsymbol{q}, \mu_m)$, which is realized in presence of the gradients $\nabla \mu_m$ and $\nabla T$, can be derived by the Boltzmann transport equation

\begin{equation}\label{EQ:Boltz_trans}
\frac{\partial g}{\partial t} + \nabla_{\boldsymbol{q}} g \cdot \frac{\partial \boldsymbol{q}}{\partial t} + \nabla_{\boldsymbol{r}}  g \cdot \frac{\partial \boldsymbol{r}}{\partial t}  = \left.\frac{\partial g}{\partial t}\right|_{collisions}.
\end{equation}

\noindent In the most general case, to describe both stationary and non stationary states, the collision term at the right hand side of Eq.(\ref{EQ:Boltz_trans}) should take into account both magnon conserving and non-conserving events. 

In the present Section we are interested to derive the behavior under stationary states, therefore we limit to describe the relaxation processes in which the value of the chemical potential $\mu_m$ of Eq.(\ref{EQ:BEQE}) is constant in time. We can evaluate the right hand side of Eq.(\ref{EQ:Boltz_trans}), in the relaxation time approximation, as

\begin{equation}\label{EQ:MC_collision}
\left.\frac{\partial g}{\partial t}\right|_{collisions} = - \frac{g(\boldsymbol{q},\mu_m) - g_{0}(\boldsymbol{q},\mu_m)}{\tau_c},
\end{equation}

\noindent where $\tau_c$ is the relaxation time associated to the collision events conserving the number of magnons and $g_0$ is the local distribution under constrained $\mu_m$. Eq.(\ref{EQ:Boltz_trans}) with the relaxation time approximation of Eq.(\ref{EQ:MC_collision}) describes how fast the non equilibrium distribution $g(\boldsymbol{q},\mu_m)$ decays into $g_{0}(\boldsymbol{q},\mu_m)$. 

It is worth to notice that from Eq.(\ref{EQ:Boltz_trans}) it is also possible to derive Eq.(\ref{EQ:f0rate2}), describing the relaxation from $g_{0}(\boldsymbol{q},\mu_m)$ to $g_{0}(\boldsymbol{q},0)$, as a special case. Indeed, in that case one fixes the distribution shape $g=g_0$ of  Eq.(\ref{EQ:BEQE}) and let the chemical potential $\mu_m$ to change in time. However, as the distribution shape is already the equilibrium one $g_0$, then a collision term like Eq.(\ref{EQ:MC_collision}) would have no effect. This is expected, because microscopic events contributing to the relaxation from $g_{0}(\boldsymbol{q},\mu_m)$ to $g_{0}(\boldsymbol{q},0)$ are all events which do not conserve the number of magnons. Then the relaxation must be described by introducing a different term, as it has been done in Section~\ref{SUBSEC:relaxation_thermal_magnons}, characterized by a different relaxation time $\tau_r$. By following this route one obtains also Eq.(\ref{EQ:f0rate2}) as a special limit of the general Eq.(\ref{EQ:Boltz_trans}).

To solve the kinetic Boltzmann equation (\ref{EQ:Boltz_trans}) for stationary states we put ${\partial g}/{\partial t}=0$. The second term at the left hand side of Eq.(\ref{EQ:Boltz_trans}) is evaluated by introducing the crystal momentum of magnons $\boldsymbol{p} = \hslash \boldsymbol{q}$ and by evaluating the associated force $\boldsymbol{F}=\partial \boldsymbol{p}/\partial t = (1/\hslash)\partial \boldsymbol{q}/\partial t $ as the gradient of the energy $\boldsymbol{F} = -\nabla_{\boldsymbol{r}} \epsilon$.  The result is $ \nabla_{\boldsymbol{q}} g \cdot {\partial \boldsymbol{q}}/{\partial t} = - \boldsymbol{v} \cdot \nabla_{\boldsymbol{r}} \epsilon \,\, {\partial g}/{\partial \epsilon}$, where $\boldsymbol{v} = ({1}/{\hslash}) \nabla_{\boldsymbol{q}} \epsilon$ is the magnon group velocity. The third term at the left hand side is evaluated by using $\boldsymbol{v} = {\partial \boldsymbol{r}}/{\partial t}$.  The resulting Boltzmann transport equation for magnons reads

\begin{equation}\label{EQ:Boltz}
g = g_0 - \tau_c \, \boldsymbol{v} \cdot \left( \nabla_{\boldsymbol{r}} g - \nabla_{\boldsymbol{r}} \epsilon \,\, \frac{\partial g}{\partial \epsilon}  \right).
\end{equation}

\noindent If the gradients are not so large, the second term at the right hand side is small and, in a first approximation, it can be evaluated at $g=g_0$ with $g_0$ given by Eq.(\ref{EQ:BEQE}). After a few passages we find the solution for the non equilibrium distribution of magnons in terms of the gradients of the intensive variables $T$ and $\mu_m$:

\begin{equation}
\label{EQ:Boltz_sol}
g  \simeq g_{0} + \tau_c \, \boldsymbol{v} \cdot \left[ \nabla_{\boldsymbol{r}} \mu_m + (\epsilon - \mu_m)\frac{\nabla_{\boldsymbol{r}} T}{T} \right] \frac{\partial g_{0}}{\partial \epsilon}.
\end{equation}

\noindent For the previous solution we observe that the effect of the second term at the right hand side is to  slightly change the shape of $g_0$ without changing the integral. Therefore the collision time $\tau_c$ can be interpreted as describing processes in which the number of magnons does not change.

%%%%%%%%%%%%%%%%%%%%%%%%%%%%%%%%%%%%%%%
%%%%%%%%%%%%%%%%%%%%%%%%%%%%%%%%%%%%%%%
\subsection{Transport coefficients}\label{SUBSEC:transp_coeff}

The coefficients of the constitutive equations (\ref{EQ:M_curr}) and (\ref{EQ:q_curr}) can be calculated by expressing the magnetization and the heat currents in terms of the non equilibrium distribution of magnons $g(\boldsymbol{q})$ given by the solution of the Boltzmann transport equation (\ref{EQ:Boltz})\cite{Wilson-1953,Solyom-2007}. The two expressions, in vector form, are

\begin{equation}
\label{EQ:jMdef}
\boldsymbol{j}_M = - 2\mu_B \boldsymbol{j}_n = - 2\mu_B \frac{1}{(2\pi)^3}\int_{\Sigma} \boldsymbol{v} \, g(\boldsymbol{q}) d^3q
\end{equation}

\begin{equation}
\label{EQ:jqdef}
\boldsymbol{j}_q = \boldsymbol{j}_u - \mu_m \boldsymbol{j}_n= \frac{1}{(2\pi)^3}\int_{\Sigma} \boldsymbol{v} \, (\epsilon-\mu_m) \, g(\boldsymbol{q}) d^3q.
\end{equation}

\noindent By inserting the approximated solution Eq.(\ref{EQ:Boltz_sol}) into the integrals Eqs.(\ref{EQ:jMdef}) and (\ref{EQ:jqdef}), we observe that only the second term of Eq.(\ref{EQ:Boltz_sol}), containing the derivative of the Bose-Einstein distribution $g_0$ (Eq.(\ref{EQ:BEQE})), is non zero. Moreover, by assuming to deal with an isotropic material and by taking both currents and forces along the $x$ direction, the integrals of Eqs.(\ref{EQ:jMdef}) and (\ref{EQ:jqdef}) are simplified. Indeed, the tensor $v_i v_j$, where the index $i$ corresponds to the component of the current and the index $j$ to the component of the force, reduces to the scalar $v_x^2$. The result of the integrations can be then directly written in terms of the coefficients of the constitutive equations (\ref{EQ:M_curr}) and (\ref{EQ:q_curr}). This way, we find the following expressions for the conductivity of the magnetization current

\begin{equation}
\mu_0\sigma_M =  \tau_{c_0} m v_m^2 \frac{1}{3\pi^2} t^{3/2}C_0,
\label{EQ:sigma_Mcalc}
\end{equation}

\noindent for the absolute thermo-magnetic power coefficient

\begin{equation}
\epsilon_M = - \frac{k_B}{2\mu_B} \frac{C_1}{C_0},
\label{EQ:epsilon_Mcalc}
\end{equation}

\noindent and for the thermal conductivity due to the magnon diffusion \cite{Douglass-1963, Sanders-1977}

\begin{equation}
\kappa = nk_B \tau_{c_0} v_m^2 \frac{1}{3\pi^2}  t^{5/2} \frac{C_0C_2-C_1^2}{C_0}.
\label{EQ:kappacalc}
\end{equation}

\noindent In the previous equations the parameter $v_m= ({k_BT_m}/{m^*})^{1/2}$ gives the scale of the thermal velocity of the spin waves, $m^* = \hslash^2/(2D)$ is the magnon effective mass and $C_r$ are coefficients defined in Appendix~\ref{APPENDIX_A}, Eq.(\ref{EQ:Cr_def}). Since we are considering the possibility that the collision time $\tau_c$ may depend on the magnon wavenumber modulus $q$ as $\tau_c = \tau_{c_0} f_c(q)$, where $f_c(q)$ is a normalized wavenumber dependence, equations (\ref{EQ:sigma_Mcalc}) and (\ref{EQ:kappacalc}) contain  explicitly the constant $\tau_{c_0}$, while $f_c(q)$ is included in the $C_r$ definition.

%%%%%%%%%%%%%%%%%%%%%%%%%%%%%%%%
%%%%%%%%%%%%%%%%%%%%%%%%%%%%%%%%
%%%%%%%%%%%%%%%%%%%%%%%%%%%%%%%%
\section{Results}\label{SEC:results}

The result of the magnon transport theory, developed in Sections~\ref{SEC:statistics_magnons} and \ref{SEC:transport_magnons}, is represented by Eqs.(\ref{EQ:tau_Mcalc}) and (\ref{EQ:sigma_Mcalc}-\ref{EQ:kappacalc}), expressing the temperature and magnetic field dependences of the phenomenological time constant $\tau_M$ appearing in the continuity equation (\ref{EQ:cont}) and of the coefficients $\sigma_M$, $\epsilon_M$ and $\kappa$ of the constitutive equations (\ref{EQ:M_curr}) and (\ref{EQ:q_curr}). Eqs.(\ref{EQ:tau_Mcalc}) and (\ref{EQ:sigma_Mcalc}-\ref{EQ:kappacalc}) depend on the coefficients $A_0^{\prime}$ and $C_r$, defined by integrals over the space $\Sigma$ of the admissible wavenumbers $\boldsymbol{q}$ that are calculated by standard methods (see Appendix~\ref{APPENDIX_A}) \cite{Wilson-1953}.

\begin{figure}[htb]
\centering
\includegraphics[width=8.8cm]{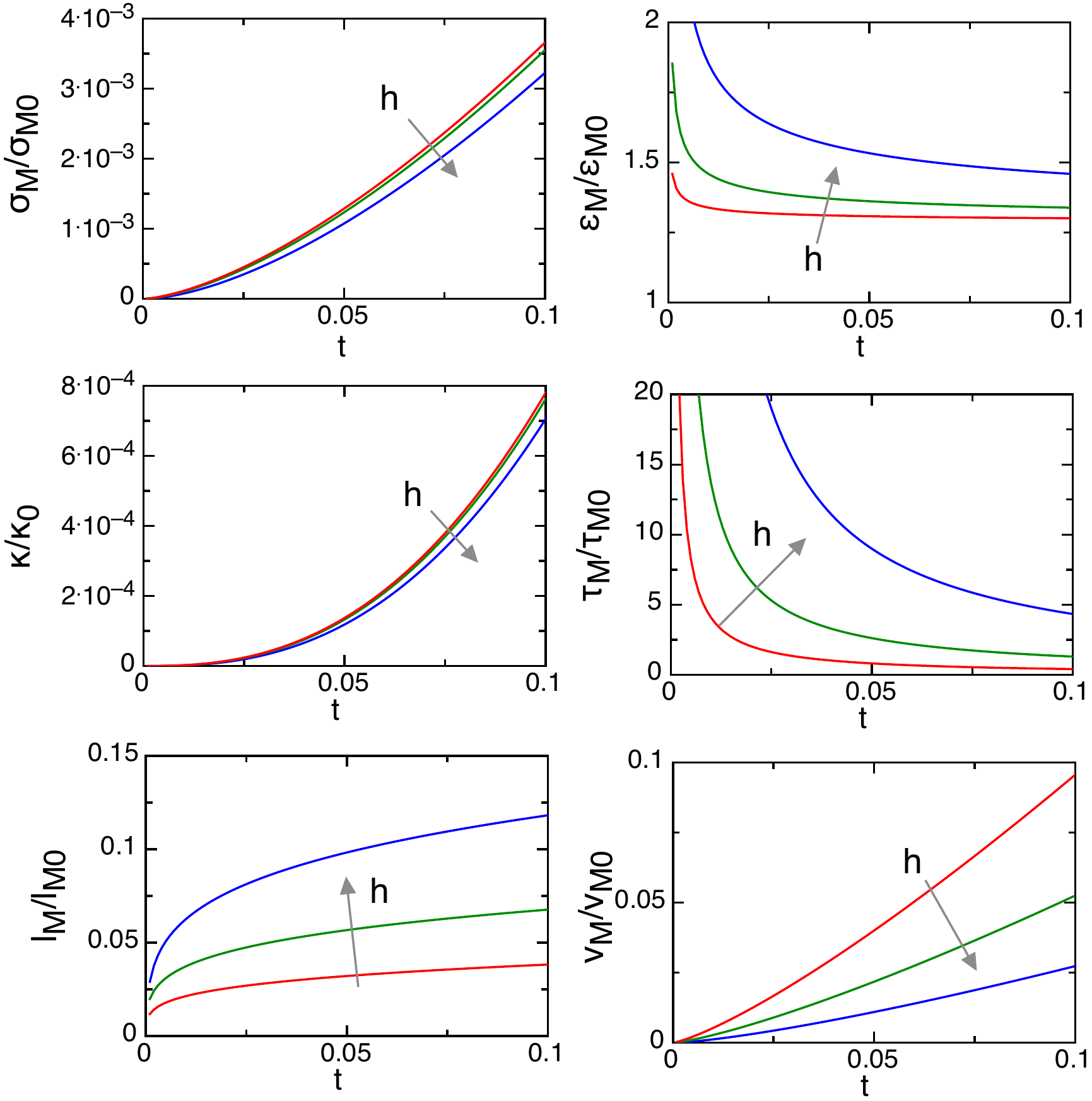}
\caption{Thermo-magnetic coefficients computed from the magnon transport theory (low temperature approximation). The coefficients are normalized to: $\sigma_{M_0}=\tau_c m v_m^2/\mu_0$, $\epsilon_{M_0}= - k_B/(2\mu_B)$, $\kappa_0=nk_B \tau_c v_m^2$, $\tau_{M_0} = \tau_r/m $, $l_{M_0} = (\tau_c\tau_r)^{1/2} v_m$, $v_{M_0} = (\tau_c/\tau_r)^{1/2} m v_m$. For all graphs, the values of $h$ are $10^{-5}$, $10^{-4}$ and $10^{-3}$ (red, green and blue, respectively).} \label{FIG:coeffLT}
\end{figure}

\begin{figure}[htb]
\centering
\includegraphics[width=8cm]{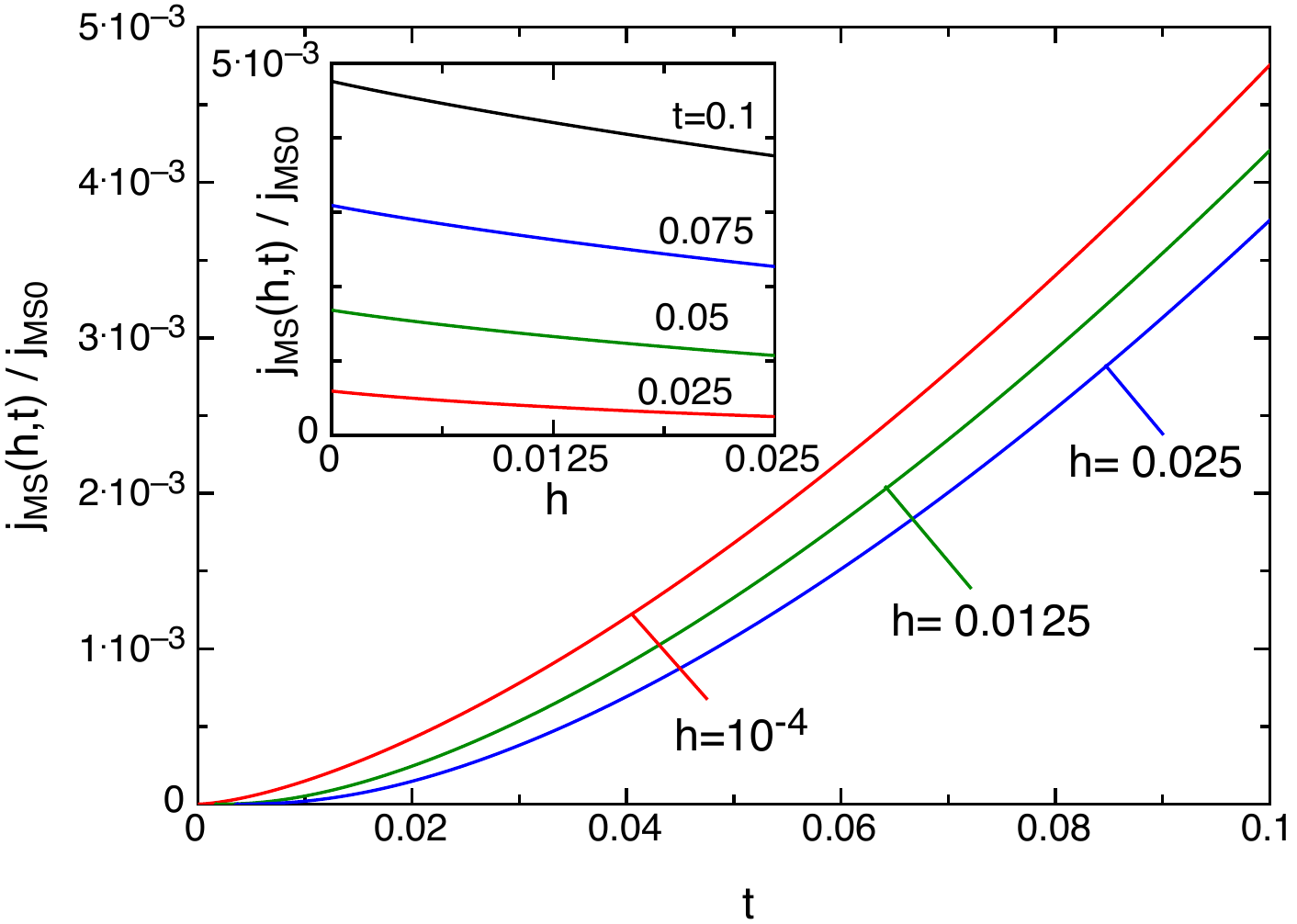}
\caption{Spin Seebeck effect for the magnon transport theory (low temperature approximation). $j_{MS}$ is the magnetization current source normalized to ${j}_{MS_0} = -\epsilon_{M_0}\sigma_{M_0}\nabla T$. The main plot shows the temperature dependence at different fields. The inset shows the field dependence at different temperatures.} \label{FIG:j0LT}
\end{figure}

At low temperature, i.e. $t\ll 1$, the coefficients can be computed by using the low $q$ approximation given by Eq.(\ref{EQ:dispersion_lq}) and the resulting analytical expressions for constant $\tau_c$ are reported in Appendices~\ref{APPENDIX_A} and \ref{APPENDIX_B} and shown in Fig.\ref{FIG:coeffLT}. The leading order dependence of the coefficients on the temperature $t$ and on the field $h$ is shown in Table \ref{TAB:1} for $t\gg h$ and $t\ll h$. As a main result we find that the absolute thermo-magnetic power coefficient $\epsilon_M$, in the limit $h\rightarrow 0$, is independent of $\tau_c$ and $t$ and its value is $\epsilon_M = -({5}/{2})(\zeta(5/2)/\zeta(3/2))k_B/(2\mu_B)$, where $\zeta(\cdot)$ is the Riemann zeta function, so that $\epsilon_M = -0.6419 \, k_B/\mu_B$. By using $k_B/\mu_B = 1.49$ K T$^{-1}$ we obtain $\epsilon_M = -0.956$ K T$^{-1}$. This result shows that the magnon accumulation in terms of $\nabla \mu_m$ is directly related to the temperature gradient $\nabla T$. Fig.\ref{FIG:coeffLT} also shows the behavior of the diffusion length $l_M$ and of the ratio $v_M = l_M/\tau_M$ entering into the expression of the current which is transmitted into the sensing layer \cite{Basso-2016}. The magnetization current source of the spin Seebeck effect $j_{MS}= - \epsilon_M \sigma_M \, \nabla T$ is shown in Fig.\ref{FIG:j0LT}. It depends on the conductivity $\sigma_M$ which is determined by the scattering mechanism. The theory predicts a decrease of the current $j_{MS}$ as a function of the field $h$, which is linear for low fields (see Fig.\ref{FIG:j0LT} and Table \ref{TAB:1}). This is an effect that was also observed in YIG/Pt spin Seebeck experiments, see Ref. [\onlinecite{Kikkawa-2015}], and it will be discussed in Section~\ref{SEC:SSE}.

\begin{figure}[htb]
\centering
\includegraphics[width=8.8cm]{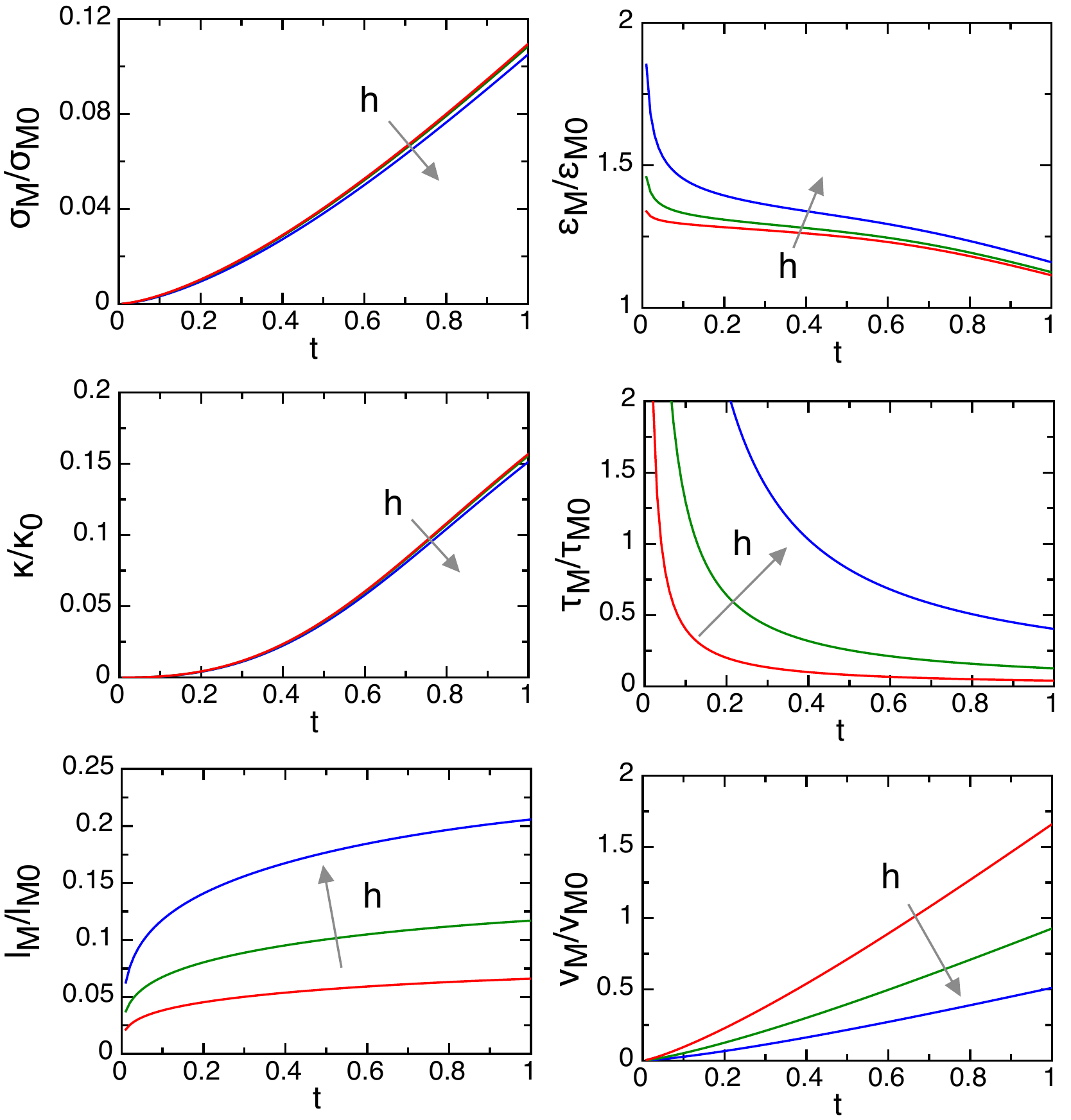}
\caption{Thermo-magnetic coefficients computed from the magnon transport theory (high temperature approximation). The coefficients are normalized to the same values reported in the caption of Fig.\ref{FIG:coeffLT}. For all graphs, the values of $h$ are $10^{-5}$, $10^{-4}$ and $10^{-3}$ (red, green and blue, respectively).} \label{FIG:coeffHT}
\end{figure}

\begin{figure}[htb]
\centering
\includegraphics[width=8cm]{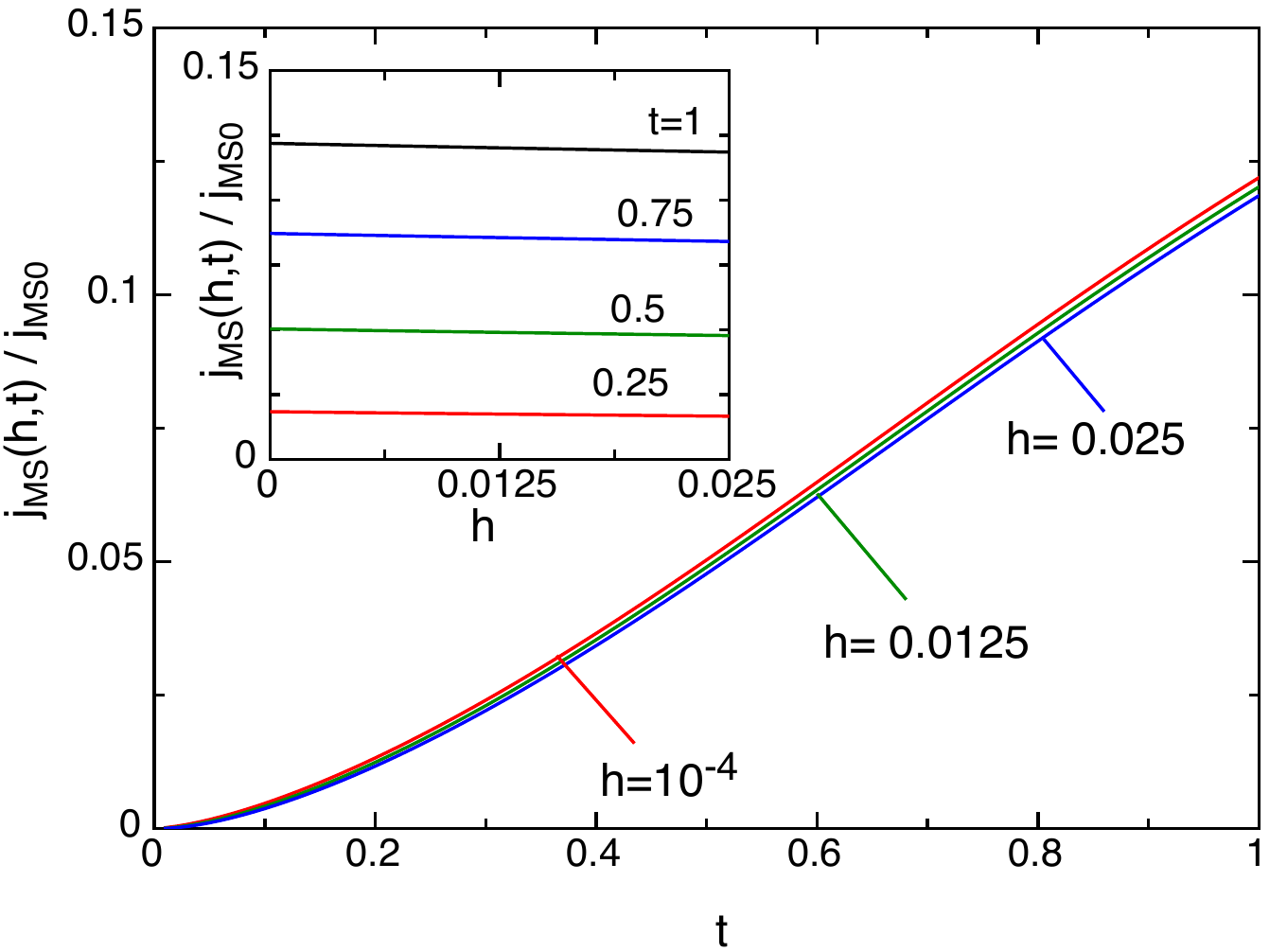}
\caption{Spin Seebeck effect for the magnon transport theory (high temperature approximation). $j_{MS}$ is the magnetization current source normalized as in Fig.\ref{FIG:j0LT}. Main plot and inset have the same meaning as in Fig.\ref{FIG:j0LT}.} \label{FIG:j0HT}
\end{figure}

At high temperature, also the high wavenumbers of the spin waves are excited and the dispersion relation of Eq.(\ref{EQ:dispersion_lq}) shall be modified in order to take into account that at high $q$ the magnon energy reaches the maximum energy level $\epsilon=\epsilon_m+\epsilon_H$. Considering the isotropic expression given in Appendix~\ref{APPENDIX_A}, the transport coefficients can be computed in the full temperature range up to $T = T_m$, i.e. $t=1$. The integrals are performed by numerical methods. Fig.\ref{FIG:coeffHT} shows the temperature $t$ and the field $h$ dependences of the thermo-magnetic coefficients, while Fig.\ref{FIG:j0HT} shows the current source $j_{MS}$ as a function of the field $h$. The behavior of the coefficients at high $t$, obtained by taking a constant $\tau_c$, does not present novel or intriguing features with respect to the low temperature case. The interesting physics of the high $t$ behavior it is rather to be searched in the wavenumber dependence of the collision time for magnons $\tau_c(q)$. This aspect will be briefly discussed in Section~\ref{SEC:conclusions}.

Another temperature region that can be explored by the theory is the very low temperature region $t\ll h$. The results are shown in Table \ref{TAB:1}. While most of the quantities goes to zero exponentially as soon as $t\rightarrow 0$, it is worth to observe the special behavior of $\epsilon_M$ and $l_M$ which do not contain the exponential decay. The consequence of this fact can be appreciated by computing the spin Peltier coefficient $\Pi_M$, which is the proportionality coefficient between the heat and the magnetization currents, $j_q  = \Pi_M j_M$. It can be derived from the constitutive equations (\ref{EQ:M_curr}) and (\ref{EQ:q_curr}) by imposing a constrained magnetization current and the condition $\nabla T = 0$. It results that $\Pi_M=\epsilon_M T$ and therefore, in the low $q$ approximation, we have

\beq
\Pi_M = - \frac{k_B}{2\mu_B} \frac{C_1}{C_0} T.
\eeq 

\noindent In the limit of very low temperature $t \ll h$ we know that $C_1/C_0 = 5/2+h/t$ (see Table \ref{TAB:1}) and thus

\beq
\Pi_M = - \left(\frac{5}{2}\frac{k_BT}{2\mu_B} + \mu_0H_0\right). 
\eeq 

\noindent We see that as soon as the temperature approaches zero, $T\rightarrow 0$, the coefficient $\Pi_M$ tends to the finite value $\Pi_M(0) = -\mu_0H_0$. This indicates the possibility to employ the spin Peltier effect as an efficient cooling mechanism close to absolute zero. However, the low temperature efficiency of spin Peltier cooling is limited by the diffusion length $l_M$ of the ferromagnet which is going to zero for $T\rightarrow 0$ as $l_M = l_{M_0}t^{1/2}$, with the normalization coefficient $l_{M_0}$ defined in the caption of Fig.\ref{FIG:coeffLT}. Since $l_M$ is the typical length over which the magnetization current can develop, the cooled region will also shrink to zero in agreement with the third principle of thermodynamics \cite{Callen-1985}. However, the slow decrease of $l_M$ as the square root of the temperature gives the possibility to explore experimentally this effect in the cryogenic region.

\begin{table*}
\centering
{\renewcommand\arraystretch{1.5}
\begin{tabular}{|c|c|c|c|}
\hline
& & \raisebox{-.5ex}{limit $t\ll h$} & \raisebox{-.5ex}{limit $t\gg h$}\\[1.5ex]
\hline
\raisebox{-.5ex}{$\sigma_M/\sigma_{M_0}$}
& \raisebox{-.5ex}{$\frac{1}{3\pi^2} t^{3/2}C_0$}
& \raisebox{-.5ex}{$\frac{1}{4\pi^{3/2}} t^{3/2} \exp\left(-\frac{h}{t}\right)$}
& \raisebox{-.5ex}{$\frac{\zeta(3/2)}{4\pi^{3/2}} t^{3/2} \left(1-\frac{2\pi^{1/2}}{\zeta(3/2)} {\left(\frac{h}{t}\right)}^{1/2}\right)$} \\[1.5ex]
\hline
\raisebox{-.5ex}{$j_{MS}/j_{MS_0}$}
& \raisebox{-.5ex}{$\frac{1}{3\pi^2} t^{3/2}C_1$}	
& \raisebox{-.5ex}{$\frac{1}{4\pi^{3/2}} t^{3/2} \left( \frac{5}{2} + \frac{h}{t}\right) \exp\left(-\frac{h}{t}\right)$}
& \raisebox{-.5ex}{$\frac{5\zeta(5/2)}{8\pi^{3/2}} t^{3/2} \left( 1 - \frac{3\zeta(3/2)}{5 \zeta(5/2)} \left( \frac{h}{t}\right) \right)$}\\[1.5ex]
\hline
\raisebox{-.5ex}{$\epsilon_M/\epsilon_{M_0}$}
& \raisebox{-.5ex}{$\frac{C_1}{C_0}$}
& \raisebox{-.5ex}{$ \frac{5}{2} + \frac{h}{t} $}
& \raisebox{-.5ex}{$\frac{5}{2}\frac{\zeta(5/2)}{\zeta(3/2)} \left( 1+\frac{2\pi^{1/2}}{\zeta(3/2)} \left( \frac{h}{t}\right)^{1/2}\right)$}\\[1.5ex]
\hline
\raisebox{-.5ex}{$\tau_M/\tau_{M_0}$}
& \raisebox{-.5ex}{$2\pi^2 t^{-1/2} \left(\frac{1}{-A_{0}^{\prime}}\right)$}
& \raisebox{-.5ex}{$4\pi^{3/2} t^{-1/2} \exp\left(\frac{h}{t}\right)$}
& \raisebox{-.5ex}{$4\pi \, t^{-1} h^{1/2}$}\\[1.5ex]
\hline
\raisebox{-.5ex}{$l_M/l_{M_0}$}
& \raisebox{-.5ex}{${\left[\frac{2}{3} t \left(\frac{C_0}{-A_{0}^{\prime}}\right)\right]}^{1/2}$}
& \raisebox{-.5ex}{$t^{1/2}$}
& \raisebox{-.5ex}{$\frac{\zeta(3/2)^{1/2}}{\pi^{1/4}}\, t^{1/4} h^{1/4}$}\\[1.5ex]
\hline
\end{tabular}}
\caption{Expressions and leading order dependence of the relevant coefficients of magnon transport theory for $t\ll h$ and $t \gg h$. The coefficients are normalized as in Fig.\ref{FIG:coeffLT} and Fig.\ref{FIG:j0LT}.}
\label{TAB:1}
\end{table*}

%%%%%%%%%%%%%%%%%%%%%%%%%%%%%%%%
%%%%%%%%%%%%%%%%%%%%%%%%%%%%%%%%
%%%%%%%%%%%%%%%%%%%%%%%%%%%%%%%%
\section{Application to the spin Seebeck effect}\label{SEC:SSE}

The knowledge of the kinetic coefficients allows to compute the spin current in a ferromagnet and to predict the amplitude of the spin Seebeck effect. Problems related to the passage of magnetization and heat currents between two adjacent media, including the spin Seebeck and the spin Peltier effects in YIG/Pt layers, can be well described within the Johnson and Silsbee approach \cite{Johnson-1987, Basso-2016}. 

To learn about the values of the parameters obtained in the theory, we need to insert typical values for the materials of interest. YIG is the most interesting material for the spin Seebeck effect and even if YIG is a ferrimagnet with a dispersion relation of several branches \cite{Cherepanov-1993}, we can apply the theory at low temperature where only the first branch is relevant. With a saturation magnetization of $M_0 = 1.96\cdot 10^{5}$ A/m \cite{Stancil-2009} and a spin wave exchange stiffness of $D = 8.6 \cdot 10^{-40}$ Jm$^{2}$ \cite{Cherepanov-1993}, we have a typical temperature $T_m = 480$ K and an effective mass $m^*= 2.5\cdot 10^{-28}$ kg, which is about $280$ times the electron mass (here we have used the relation $M_0 = \mu_B/a^3$ with $a \sim 0.36\cdot 10^{-9}$ m being the typical distance between magnetic ions). The field $H_0$, appearing in the ground state energy $\epsilon_H$ of Eq.(\ref{EQ:dispersion_lq}), is the sum of the anisotropy field along one of the easy axis $\mu_0H_{AN} \sim 0.017$ T \cite{Stancil-2009} and of the applied field $H_a$. For low applied fields, $H_a \ll H_{AN}$, we have $h \sim 0.5\cdot10^{-4}$, while for high applied fields, $H_a \gg H_{AN}$, we have $h = 10^{-3}$ at $\mu_0H_a = 0.35$ T. The remaining parameters are $m = 6.8\cdot10^{-4}$ and $v_m = 6.6 \cdot 10^{3}$ ms$^{-1}$. By assuming $\tau_r \approx \tau_c=1$ ns, we obtain $j_{MS_0}/\nabla T = k_B\tau_c m v_m^2/(2\mu_0\mu_B) = 17.6$ A s$^{-1}$ K$^{-1}$ m. 

The previous values can be used to compute the order of magnitude of the spin Seebeck effect, by evaluating the magnetization current which is injected into the Pt layer by means of the properties of the inverse spin Hall effect in Pt. The injection of the magnetization current source into the metallic layer is due to the conversion of the magnetic moment of the magnons at the interface into a spin accumulation in Pt. The microscopic mechanisms involved \cite{Brataas-2000, Ohe-2011, Qiu-2013, Qiu-2013, Hoffmann-2013, Tveten-2015, Chotorlishvili-2015, Bender-2015} as well as the role played by imperfect interfaces \cite{Qiu-2015} have been already analyzed in detail. From the point of view of the thermodynamic theory of Johnson and Silsbee \cite{Johnson-1987} the passage of the magnetization current between two layers is described by solving the diffusion problem for the two materials and by imposing the appropriate boundary conditions \cite{Basso-2016}. Each layer is characterized by a diffusion length for the magnetization current $l_M = (\mu_0 \sigma_M \tau_M)^{1/2}$ which is related to intrinsic properties of each material: the magnetization conductivity $\sigma_M$ and the time constant $\tau_M$. At the interface between the two layers, one requires the magnetization current density $j_M$ to be continuous and the potentials $H^*$ to be related by the interface resistance \cite{Johnson-1987}. 

In the present paper we are interested in the evaluation of the order of magnitude of the spin Seebeck effect that can be obtained in bulk YIG at low temperature. Using the inverse spin Hall effect of Pt, the magnetization current is converted into an electric voltage with a coefficient $1.3\cdot10^{-4}$ Vm$^{-1}$A$^{-1}$ s \cite{Basso-2016}. Therefore we find that the spin Seebeck signal at $T=48$ K ($t=0.104$) is 3.7 $\mu$V/K at low fields and it decreases to 2.9 $\mu$V/K at 9 T ($h=0.025$) in agreement with the measurements reported in Ref. [\onlinecite{Kikkawa-2015}] for a YIG slab of 1 mm of thickness and a thin Pt layer \cite{Basso-2016}. In the theory, this effect is the consequence of the change of the Bose-Einstein distribution, Eq.(\ref{EQ:BEEQ}), with the field dependent energy term $\epsilon_H$, which is strongly pronounced for low $q$ magnons. Similar conclusions have been achieved in Refs.[\onlinecite{Rezende-2014}], [\onlinecite{Ritzmann-2015}] and [\onlinecite{Diniz-2016}]. A detailed comparison with experimental data at several temperatures and fields requires also to take care of the current injection at the YIG/Pt interface, which will be the aim of future works. The agreement found at 48 K is obtained by using a transmission coefficient for the current at the YIG/Pt interface of 75\% which is quite close to the value of 50\% that was used at room temperature in Ref. [\onlinecite{Basso-2016}].

%%%%%%%%%%%%%%%%%%%%%%%%%%%%%%%%
%%%%%%%%%%%%%%%%%%%%%%%%%%%%%%%%
%%%%%%%%%%%%%%%%%%%%%%%%%%%%%%%%
\section{Discussion and conclusions}\label{SEC:conclusions}

We have used the Boltzmann transport theory in the relaxation time approximation to describe the thermal transport of spin waves in an insulating ferromagnet in the framework of the Johnson and Silsbee thermodynamic theory \cite{Johnson-1987}. Within this approach, the thermodynamic force associated with the magnetization current is the potential $H^*$ which, in the case of magnon transport, is directly related to the magnon chemical potential, that is $ \mu_m  = 2 \mu_0\mu_BH^*$. The kinetic coefficients of the constitutive thermo-magnetic transport equations are computed analytically by assuming an isotropic dispersion relation for magnons. The main outcome of the study is that the transport coefficients are analytical expressions containing the microscopic parameters of the material.

We have found several interesting properties in the magnetic field and temperature dependences of the kinetic coefficients. The absolute thermo-magnetic power coefficient $\epsilon_M$, relating the gradient of the potential of the magnetization current and the gradient of the temperature, in the limit of low temperature and low field, is a constant $\epsilon_M = -0.6419 \, k_B/\mu_B$. In the limit of very low temperatures the spin Peltier coefficient $\Pi_M$, relating the heat and the magnetization currents, tends to a finite value which depends only on the amplitude of the magnetic field. This indicates the possibility to exploit the spin Peltier effect as an efficient cooling mechanism in cryogenics.

By inserting the typical values for YIG, one of the most studied spin Seebeck materials, we have computed the order of magnitude of the spin Seebeck experiments in bulk YIG. It is worth noting that while the theory is able to reproduce rather well the initial increase of the spin Seebeck effect as a function of temperature up to around 50 K, as reported in Ref. [\onlinecite{Kikkawa-2015}], the predictions above 50 K are much higher than the measured ones. This difference deserves special attention. Considering YIG, there are a certain number of important aspects that have been neglected here: the presence of different branches which are active above the 40 K region and the special dispersion relation of ferrimagnets \cite{Cherepanov-1993}. Another important point to mention is the fact that, as soon as the temperature is increased, the saturation magnetization $M_s$ decreases significantly below the zero temperature value $M_0$ and this effect is not only caused by thermal spin waves. At intermediate temperatures the equilibrium magnetization $M_s(T)$ is mainly ascribed to localized magnetic fluctuations and the thermal spin waves, that can build up upon $M_s$, play only a minor role in determining the thermal content of the system. A theory taking into account the superposition of both local fluctuations and spin waves may be able to describe the details of the room temperature spin Seebeck effect. An interesting development along this line can be envisaged in considering an explicit dependence of the scattering time constant $\tau_c(q)$ on the magnon wavenumber $q$, as in Ref. [\onlinecite{Rezende-2014}] and  [\onlinecite{Diniz-2016}]. Indeed, the analytical results of the present theory can be extended to the case of a $\tau_c(q)$ expressed as a power series in $q$. This can be the subject of a future development of the present work.

%%%%%%%%%%%%%%%%%%%%%%%%%%%%%%
%%%%%%%%%%%%%%%%%%%%%%%%%%%%%%
%%%%%%%%%%%%%%%%%%%%%%%%%%%%%%
\section*{Acknowledgments}
This work has been carried out within the Joint Research Project EXL04 (SpinCal), funded by the European Metrology Research Programme. The EMRP is jointly funded by the EMRP participating countries within EURAMET and the European Union.

%%%%%%%%%%%%%%%%%%%%%
%%%%%%%%%%%%%%%%%%%%%
%%%%%%%%%%%%%%%%%%%%%
\appendix

%%%%%%%%%%%%%%%%%%%%%%%%%%%%%%%%
%%%%%%%%%%%%%%%%%%%%%%%%%%%%%%%%
%%%%%%%%%%%%%%%%%%%%%%%%%%%%%%%%
\section{Calculation of coefficients}\label{APPENDIX_A}

%%%%%%%%%%%%%%%%%%%%%%%%%%%%%%%%
%%%%%%%%%%%%%%%%%%%%%%%%%%%%%%%%
\subsection{Definitions of the coefficients}\label{SUBAPPENDIX:coeff_def}

The equilibrium and transport coefficients entering the theory are here defined as
\beq
A_{r} = \frac{(2\pi)^2(k_BT_m)^{3/2}}{n(k_BT)^{r+3/2}} \frac{1}{(2\pi)^3} \int_{\Sigma} \epsilon^{r} g_0 d^3q,
\label{EQ:Ar_def}
\eeq

\beq
A_0^{\prime} = (2\pi)^2\frac{(k_BT_m)^{3/2}}{n (k_BT)^{1/2}} \frac{1}{(2\pi)^3}\int_{\Sigma} \frac{\partial g_{0}}{\partial \epsilon}  d^3q,
\label{EQ:A01_def}
\eeq
and
\begin{equation}
C_r \!=\! - \frac{6\pi^2 (k_BT_m)^{3/2}m^*}{n(k_BT)^{3/2+r}} \frac{1}{(2\pi)^3}\!\!\int_{\Sigma} \!f_{c}(q) v_x^2 (\epsilon-\mu_m)^r \frac{\partial g_{0}}{\partial \epsilon} d^3q
\label{EQ:Cr_def}
\end{equation}

\noindent where $\Sigma$ is the first Brillouin zone of the reciprocal space, $n$ is the density of elementary magnetic moments and $f_c(q)$ is the normalized wavenumber dependence of the collision time $\tau_c$. For the calculation of the integrals, a standard method is adopted \cite{Wilson-1953}. Since all the coefficients in Eqs.(\ref{EQ:Ar_def}-\ref{EQ:Cr_def}) contain integrals $(2\pi)^{-3}\int d^3q$ extending over $\Sigma$, their computation for a cubic lattice with lattice constant $a$ gives rise to a factor $1/a^3$ which corresponds to the number of degrees-of-freedom per unit volume of the system. Then, the shape of $\Sigma$ is replaced by a sphere $\Sigma_0$ in the $\boldsymbol{q}$ space of radius $q_0 = \pi/a_0$ giving rise to the same number of degrees-of-freedom. This corresponds to take $a_0 = ({\pi}/{6})^{1/3} a$. This approximation is exact at low temperature for any lattice. 

%%%%%%%%%%%%%%%%%%%%%%%%%%%%%%%%
%%%%%%%%%%%%%%%%%%%%%%%%%%%%%%%%
\subsection{Low $q$ approximation}\label{SUBAPPENDIX:low_q_approx}

At low temperature we take the low $q$ approximation of the dispersion relation of Eq.(\ref{EQ:dispersion_lq}). The group velocity is $ \boldsymbol{v} = (2D/\hslash) \boldsymbol{q} $ and the energy of the state $q=q_0$ is $ \epsilon_0 = \pi^{2} D/a_0^2$.  The integral over the $\Sigma_0$ sphere on the $\boldsymbol{q}$ space is first performed over the spherical angles $\theta_q$ and $\varphi_q$ giving a $4\pi$ factor. Then the integrals reduce to the form

\beq
\label{EQ:int_energy}
\frac{1}{(2\pi)^3}\int_{\Sigma_0} d^3q = \frac{n}{(2\pi)^2}\frac{1}{(k_BT_m)^{3/2}} \int (\epsilon-\epsilon_H)^{1/2} d\epsilon.
\eeq

The equilibrium coefficients $A_r$ are calculated from Eq.(\ref{EQ:Ar_def}) and Eq.(\ref{EQ:int_energy}) by changing the variable of integration to $y = (\epsilon-\epsilon_H)/(k_BT)$. We obtain 

\beq
A_{r} = \int_{0}^{y_0} \frac{y^{1/2} (y+y_H)^{r}}{\exp(y+y_H)-1} dy
\eeq

\noindent where $y_H = \epsilon_H/(k_BT)= h/t$ and $y_0 = \epsilon_0/(k_BT)$. If the temperature is low, we can let the upper limit of the integral to go to infinity, $y_0\rightarrow \infty$, and the obtained expression will be a function of $y_H$ only. In particular

\beq
A_r = \sum_{i=0}^{r} \, {{r}\choose{i}} \,  y_H^{r-i} \, I_{1/2+i}(y_H),
\eeq

\noindent where we have used the binomial coefficient $\binom{r}{i}$ and we have defined the integral

\beq
I_s(y_H) = \int_{0}^{\infty} \frac{y^{s}}{\exp(y+y_H)-1} dy
\label{EQ:Is_def}
\eeq

\noindent which can be computed analytically (see Appendix~\ref{APPENDIX_B}). For small $y_H$ we have \cite{Robinson-1951}

\beq
I_s(y_H) = \Gamma(s+1)\textrm{Li}_{s+1}(e^{-y_H})
\label{EQ:Is_result}
\eeq

\noindent where $\Gamma(\cdot)$ is the gamma function and $\textrm{Li}_n(z)$ is the polylogarithm function \cite{Lewin-1981}. As $\textrm{Li}_n(1) = \zeta(n)$, where $\zeta(\cdot)$ is the Riemann zeta function, we have

\beq
I_s(0) = \Gamma(s+1)\zeta(s+1).
\eeq

\noindent For example, the coefficient for the number of magnons is $ A_{0} = I_{1/2}(y_H)$ and the coefficient for the average energy is $ A_{1} = I_{3/2}(y_H) + I_{1/2}(y_H) y_H$. For $h=0$ and low $q$, we have $A_r(0)=\Gamma(3/2+r)\zeta(3/2+r)$.

In an analogous way we compute the coefficient $A_0^{\prime}$ given by Eq.(\ref{EQ:A01_def}). By making an integration by parts we find 

\beq
A_0^{\prime} = -\frac{1}{2} I_{-1/2}(y_H).
\eeq

\noindent The coefficient for $h=0$ is $A_0^{\prime}(0) = -({1}/{2})\Gamma(1/2)\zeta(1/2)$.

Finally, the integral of Eq.(\ref{EQ:Cr_def}) is calculated by taking a constant collision time ($f_c(q)=1$). After integration by parts we have

\begin{equation}
C_r = \int_{0}^{y_0} \, \frac{\partial \eta_r}{\partial y} g_{0} dy
\end{equation}

\noindent with $\eta_r = y^{3/2} \, (y+y_H)^r$. By writing $\partial \eta_r/\partial y$ as

\beq
\frac{\partial \eta_r}{\partial y} = \frac{3}{2}y^{1/2}(y+y_H)^r + r y^{3/2}(y+y_H)^{r-1},
\eeq

\noindent we obtain

\beq
\frac{\partial \eta_r}{\partial y} \!\!=\! \frac{3}{2}y^{1/2}\sum_{k=0}^{r} \!{{r}\choose{k}} y_H^{r-k}y^k + r y^{3/2}\sum_{k=0}^{r-1} \!{{r-1}\choose{k}} y_H^{r-1-k}y^k. 
\eeq

\noindent By using the identity ${{r}\choose{k}} = \frac{r}{k}{{r-1}\choose{k-1}}$ we have

\begin{equation}
\frac{\partial \eta_r}{\partial y} = \sum_{k=0}^{r} \left( \frac{3}{2} + k \right) \, {{r}\choose{k}} \,  y_H^{r-k} \, y^{1/2+k}.
\end{equation}

\noindent Now, by letting $y_0 \rightarrow \infty$ we obtain

\begin{equation}
C_r = \sum_{k=0}^{r} \left( \frac{3}{2} + k \right) \, {{r}\choose{k}} \,  y_H^{r-k} \, I_{1/2+k}(y_H).
\end{equation}

\noindent For $h=0$ we have $C_{r}(0)=({3}/{2}+r) \Gamma(3/2+r)\zeta(3/2+r)$.

%%%%%%%%%%%%%%%%%%%%%%%%%%%%%%%%
%%%%%%%%%%%%%%%%%%%%%%%%%%%%%%%%
\subsection{High $q$ isotropic approximation}\label{SUBAPPENDIX:high_q_approx}

\begin{figure}[htb]
\centering
\includegraphics[width=7cm]{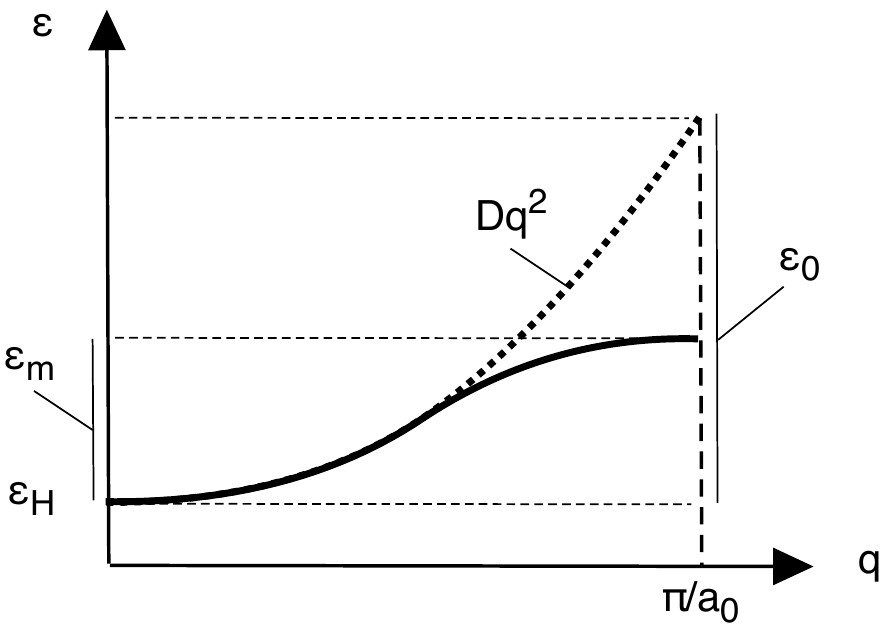}
\caption{Magnon dispersion relations. The dashed line is the low $q$ approximation of Eq.(\ref{EQ:dispersion_lq}). The full line is the high $q$ isotropic approximation of Eq.(\ref{EQ:dispersion_hq}).} \label{FIG:dispersion}
\end{figure}

At high temperature also the high wavenumbers of the spin waves are excited, so we assume an expression for the dispersion relation which is isotropic and reaches the maximum energy $\epsilon=\epsilon_m+\epsilon_H$ at $q=q_0$ (see Fig.\ref{FIG:dispersion}). An expression with this property is 

\beq
\epsilon = \epsilon_m \sin^2(a_0 q /2) + \epsilon_H
\label{EQ:dispersion_hq}
\eeq

\noindent where $\epsilon_m = 4D/a_0^2$ and the modulus of the group velocity is $v  = (2 D)/(a_0 \hslash) \sin(a_0 q)$. The isotropic high $q$ approximation would be in principle valid at any temperature, but we have to consider that the whole spin wave picture is based on the approximation of small deviations from the equilibrium state $M_0$, so in practice it is reasonable to take $T = T_m$ ($t=1$) as an upper limit. The integrals are performed by changing variable from the reciprocal space to the energy and then by using $x=(\epsilon-\epsilon_H)/\epsilon_m$. The integrals become

\beq
\frac{1}{(2\pi)^3}\int_{\Sigma_0} d^3q = \frac{12n}{\pi^3} \int_0^1 \rho_m(x) dx
\eeq

\noindent where $\rho_m(x) = [x(1-x)]^{-1/2}(\arcsin(x^{1/2}))^2$. Using $x$ as a new integration variable, we have:

\begin{align}
A_{r} &= y_m^{3/2} \int_{0}^{1} (y_mx+y_H)^r g_0 \rho_m(x) dx,\\
A_0^{\prime} &= y_m^{1/2} \int_{0}^{1}  \frac{\partial g_{0}}{\partial x} \rho_m(x) dx,
\end{align}

\noindent where $g_0 = (\exp\left(y_mx+y_H\right)-1)^{-1}$ and $y_m = \epsilon_m/(k_BT) = 4 ({6}/{\pi})^{2/3}/t$, and for the transport coefficients, by using the same assumption of the low $q$ case, we find

\beq
C_r = - y_m^{3/2} \int_{0}^{1}  x(1-x) (y_H+y_mx)^r \frac{\partial g_{0}}{\partial x} \rho_m(x) dx.
\eeq

\noindent The coefficients can be computed numerically and they can be represented as a function of the normalized temperature $t$ and of the ratio $h/t$.

%%%%%%%%%%%%%%%%%%%%%%%%%%%%%%%%
%%%%%%%%%%%%%%%%%%%%%%%%%%%%%%%%
%%%%%%%%%%%%%%%%%%%%%%%%%%%%%%%%
\section{Expressions for $I_s(y_H)$}\label{APPENDIX_B}

At low temperature the coefficients are all expressed in terms of the integral of Eq.(\ref{EQ:Is_def}) which is a function of $y_H = h/t$. We can subdivide the behavior of $I_s(y_H)$ into: \textit{a}) $y_H\ll 1$, i.e. $h \ll t$, which corresponds to the behavior at low fields and \textit{b}) $y_H\gg 1$, i.e. $t \ll h$, which corresponds to the behavior at very low temperatures. 
%%%%%%%%%%%%%%%%%%%%%%%%%%%%%%%%
\paragraph{$y_H\ll 1$.} At low $y_H$ the integral can be expressed in terms of the polylogarithm function (Eq.(\ref{EQ:Is_result})). A power expansion of the polylogarithm for small $x$ is \cite{Robinson-1951}

\beq
\textrm{Li}_n(e^x) = \Gamma(1-n)(-x)^{n-1}+\sum_{k=0}^{\infty}\frac{\zeta(n-k)}{k!}x^k.
\eeq

\noindent The expression for $I_s(y_H)$ of Eq.(\ref{EQ:Is_result}) then becomes

\beq
I_s(y_H) = \Gamma(s+1)\left[ \Gamma(-s)y_H^{s}+\sum_{k=0}^{\infty}(-1)^{k}\frac{\zeta(s+1-k)}{k!}y_H^k\right]
\label{EQ:Is_calc}
\eeq

\noindent where the power series can be computed up to the desired order. Table \ref{TAB:1} reports the expression of the relevant parameters for magnon transport theory by taking the leading order in $h/t$ only, while the curves reported in the figures are computed by using Eq.(\ref{EQ:Is_calc}) up to the order $k=9$.

%%%%%%%%%%%%%%%%%%%%%%%%%%%%%%%%
\paragraph{$y_H\gg 1$.} In the limit $y_H \rightarrow \infty$ the expression of Eq.(\ref{EQ:Is_def}) is 

\beq
I_s(y_H) = \Gamma(s+1)\exp(-y_H).
\eeq

\noindent This means that in the $T \rightarrow 0$ limit with $H_0\neq0$ all the integrals $I_s$ decay exponentially to zero.

\bibliography{00_SS}

\end{document}